# Crossing exceptional points in non-Hermitian quantum systems

Friederike U.J. Klauck,[1] Matthias Heinrich,[1] Alexander Szameit,[1] Tom A.W. Wolterink[1]*

- [1] Institute of Physics, University of Rostock, Rostock, Germany
- * tom.wolterink@uni-rostock.de

**Abstract**
Exceptional points facilitate peculiar dynamics in non-Hermitian systems. Yet, in photonics, they have mainly been studied in the classical realm. In this work, we reveal the behavior of two-photon quantum states in non-Hermitian systems across the exceptional point. We probe the lossy directional coupler with an indistinguishable two-photon input state and observe distinct changes of the quantum correlations at the output as the system undergoes spontaneous breaking of parity-time symmetry. Moreover, we demonstrate a switching in the quantum interference of photons directly at the exceptional point, where Hong-Ou-Mandel dips are transformed into peaks by a change of basis. These results show that quantum interference and exceptional points are linked in curious ways that can now be further explored.

**Introduction**
Exceptional points (EPs) are remarkable singularities occurring in non-Hermitian systems, signaling a branch point in parameter space (*1–3*). At an exceptional point, two or more eigenvalues as well as their associated eigenstates coalesce, creating fascinating dynamics. A class of non-Hermitian systems of particular interest are quantum systems obeying parity-time symmetry, since they can exhibit real eigenvalues despite their non-Hermiticity. At the exceptional point, PT-symmetric systems experience a symmetry-breaking phase transition where their eigenvalues turn complex (*4–6*). Photonics provides an excellent platform to construct non-Hermitian systems and explore exceptional points (*7*). Indeed, the evolution of classical light in non-Hermitian systems has been investigated across the exceptional point (*8–12*) and in encircling the EP (*13, 14*), showing for instance chiral behavior and swapping of eigenstates, and the concept has made its way to applications as in enhanced sensors (*15, 16*) and robust lasers (*17, 18*). Although extensive research has been performed on EPs in photonic systems, all previous experimental studies focus on first quantization systems, while disregarding the quantum nature of light itself. The interplay of exceptional points with light in second quantization, governing the behavior of indistinguishable bosons, therefore remains almost unexplored. Probing non-Hermitian systems with quantum light offers new dynamics, as the introduction of losses substantially affects the quantum statistics of light in curious ways. Especially two-photon interference changes depending on whether one observes a full Hermitian system or a non-Hermitian subsystem that experiences loss (*19–23*). Various theoretical works consider fermionic and bosonic dynamics throughout the unbroken and broken PT-symmetry phase (*24–28*). Recently, PT-symmetric quantum interference was observed for the first time (*29*), followed by the quantum simulation of coupled PT-symmetric Hamiltonians (*30*) on a photonic platform. These advances now enable experimental research of the interplay of exceptional points with quantum correlations.

In this work, we explore how an exceptional point of a non-Hermitian systems affects the quantum behavior of light. As shown in Fig. 1, we vary a single parameter of the quantum photonic system to cross the exceptional point along a one-dimensional trajectory. Probing the system with distinguishable photons reveals dynamics based on classical interference, whereas indistinguishable photons interrogate the quantum correlations in the non-Hermitian system. To uncover direct signatures of the exceptional point in a non-Hermitian bosonic system, we consider a suitable rotation of basis of the two-photon input and output states (*31*, *32*) by which the quantum interference of indistinguishable photons directly changes sign at the exceptional point. We experimentally demonstrate this behavior in a Hong-Ou-Mandel experiment (*33*), where we observe a HOM dip whenever the non-Hermitian system is in the unbroken phase with real eigenvalues, and a peak in the broken phase.

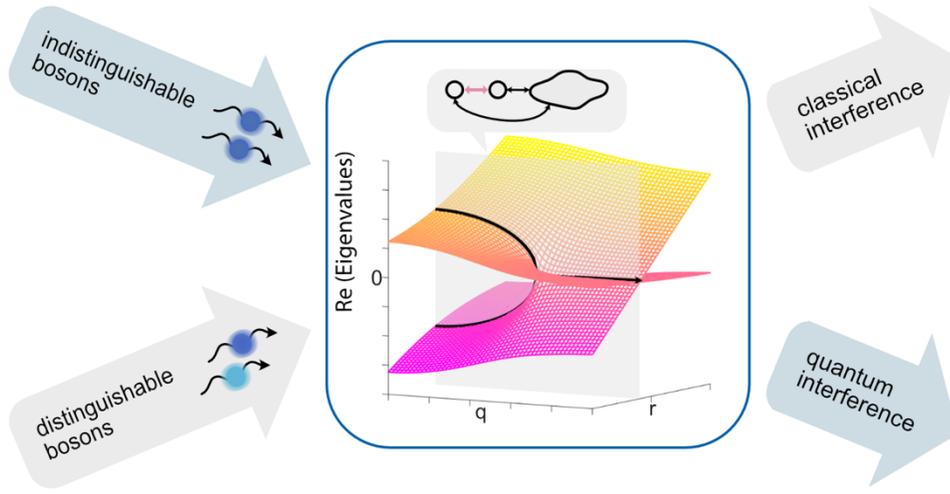

**Fig. 1. Crossing the exceptional point**. *Exceptional points occur in two-mode systems that exhibit gain and/or loss by coupling through a reservoir. To traverse the exceptional point along one coordinate q, indicated by the gray plane, a single free parameter of the system is varied. While crossing the exceptional point also affects the behavior of classical light, we probe the evolution of indistinguishable photons to observe the effects of the EP crossing on quantum interference.*

## Results

We study the dynamics of boson pairs across an exceptional point in an integrated photonic lossy directional coupler. This parity-time symmetric coupler consists of two coupled waveguides. We introduce non-Hermiticity into the system by coupling one of the modes unidirectionally to a reservoir (see Fig. 2 A), such that the two-waveguide system effectively exhibits Markovian loss. Note that the absence of gain ensures that a quantum state propagates through the system without incurring noise (*34*, *35*). It is described by the non-Hermitian Hamiltonian

$$H = \begin{pmatrix} 0 & \kappa \\ \kappa & -2i\gamma \end{pmatrix},$$

with coupling $\kappa$ and loss $\gamma$. This passive PT-symmetric system exhibits an exceptional point when the coupling of the lossy waveguide to the reservoir equals the coupling between waveguides, see Fig. 2B. As long as losses are lower than the coupling between waveguides, the system shows real eigenvalues. At the exceptional point itself, PT symmetry spontaneously breaks, and all eigenvalues become imaginary for losses exceeding coupling (*4*). Note that since the entire system is entirely passive, a linear offset to the imaginary part of the eigenvalue spectrum appears due to global damping. We study the effects of this PT-symmetry breaking transition on two-photon quantum states propagating through the system, observing only two-photon output states. The full non-Hermitian evolution of quantum states in these lossy directional couplers is analytically described in various ways (29, 36). Yet, post-selecting on the cases where no photons are lost (37), the propagation along $z$ is described by the classical linear propagator $U = e^{iHz}$, that now describes a non-unitary evolution.

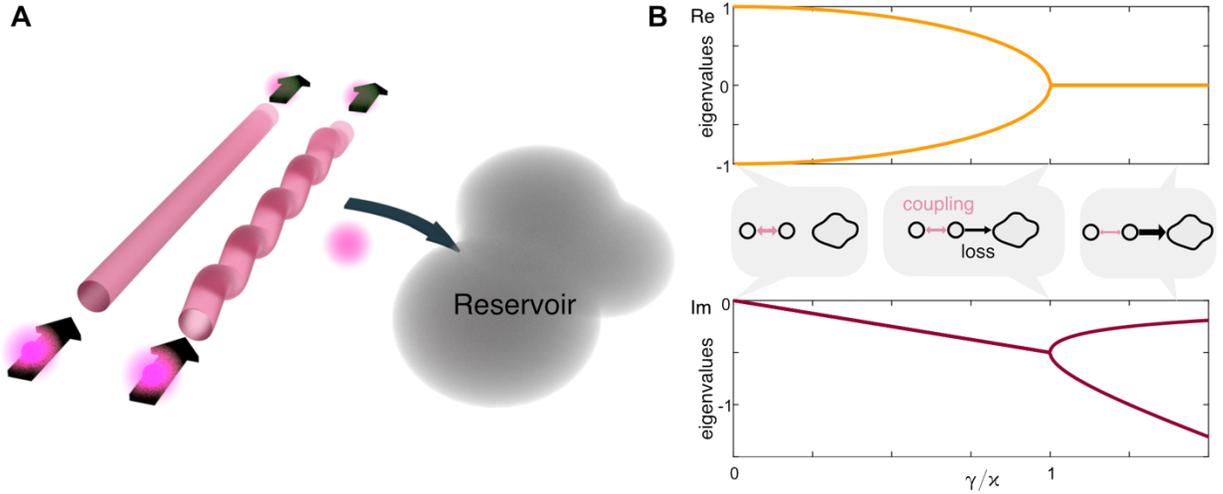

*Fig. 2. The lossy directional coupler* (A) *consists of two coupled waveguides, one of which is coupled to a reservoir to establish Markovian losses. We inject two indistinguishable photons into this system and observe all two-photon output states.* (B) *In the eigenvalue spectrum of the lossy directional coupler, the exceptional point occurs where coupling between the two modes is equal to the coupling to the reservoir. By increasing the coupling to the reservoir (i.e. the loss), the photonic system can be pushed into the broken PT-symmetry phase.*

To realise the PT breaking transition, we construct a sequence of otherwise identical couplers with increasing loss, see Fig. 3A), ranging from the lossless, Hermitian case up to above the exceptional point. In the experiment, pairs of coupled waveguides are inscribed using the femtosecond direct laser writing technique (38). For ease of handling, fan-in/fan-out sections enclose the coupler. Losses are implemented through sinusoidal undulations, where amplitude and period length of the sine control the loss in the waveguide (39). By varying the amplitude of the sine while keeping all other parameters fixed, we fabricate directional couplers of different loss values. The length of the coupling section was chosen to yield a 50/50 coupler in the lossless case. For the quantum experiment, two indistinguishable photons are generated through type I spontaneous parametric down conversion in a bismuth triborate (BiBO) crystal and collected into fibres. Changing the collection position of the photons allows to independently tune their respective time delays, and thereby their distinguishability. The photons are coupled into the waveguide chip through a fibre array with 82 µm pitch, propagate through the sample, and are collected by a second fibre array to measure all two-photon coincidences.

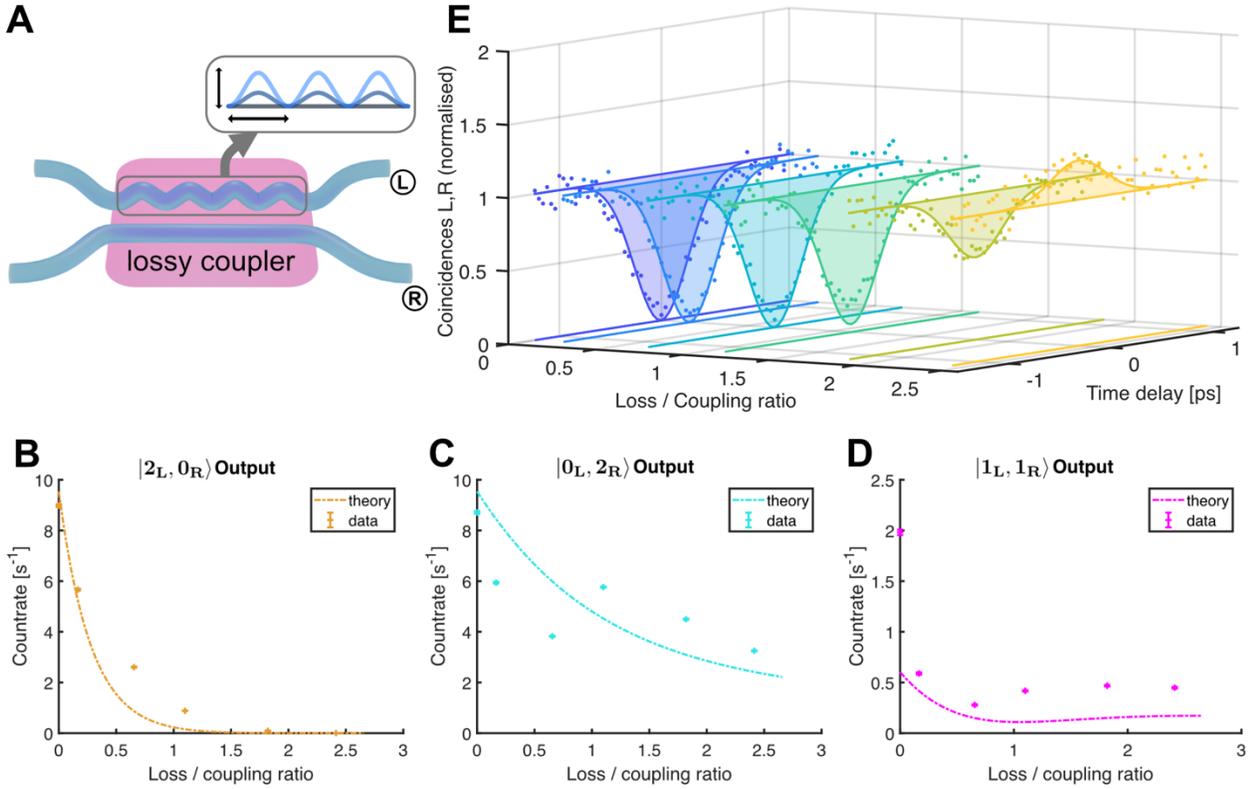

*Fig. 3. Two-photon correlations in the lossy directional coupler (A) Waveguide structure of the lossy directional coupler. We implement losses through sinusoidal undulations, whereas fan-in/fan-out sections enclose the coupler to match the pitch of the fibre array. For different structures, losses are increased through ramping up the sine amplitude, while all other parameters are kept constant. Two indistinguishable photons enter the directional coupler as $|1_L, 1_R\rangle$-input. At the output, all remaining two-photon coincidences are measured: either two photons coincide in the lossy waveguide, see (B), both photons bunch in the lossless waveguide (C), or the two photons are found in different waveguides (D). Data show error bars based on Poissonian click-statistics. (E) Hong-Ou-Mandel interference for each of the lossy directional couplers. Coincidences are normalized to the count rate of distinguishable photons. For higher losses, the probability of detecting two indistinguishable photons exceeds that for distinguishable photons, resulting in a HOM peak. The analytical theory for the experimental parameters is indicated by the lines in (B) – (E).*

**Measurements of two-photon correlations in the lossy directional coupler**

The correlation measurement performs a projection of the output state only on those events, where both photons 'survived'. Fig. 3B)-D) show the measurements of the two-photon coincidences for directional couplers of increasing loss with constant coupling length and coupling. Starting with a lossless 50/50 coupler, we observe that the photons bunch together into a single waveguide, showing Hong-Ou-Mandel interference. With increasing loss, the probability of finding two photons in the lossy waveguide decreases, as shown in Fig. 3B. The probability of the photons bunching in the lossless waveguide likewise decreases with increasing loss, but systematically exceeds that in the lossy waveguide. Meanwhile, the probability of finding a single photon in each waveguide rises. The experimental data show the same behavior as an analytical model, which considers experimental parameters and the source characteristics. The transition through and above the exceptional point at $\gamma = \kappa$ is smooth for all these observables, without apparent signature of the exceptional point. As they contain the quantum interference, we further examine the cases where photons are found in distinct waveguides. We record the two-photon coincidences $\Gamma_{1,2}$ between the two output waveguides while changing the time delay between the photons in a Hong-Ou-Mandel (HOM) type experiment (*33*), see Fig. 3E. The data points show coincidence events of the two photons in different waveguides, normalized to the count rate in the distinguishable case. The theoretical predictions of the HOM interference for the given experimental parameters are drawn as solid lines, matching the data well. We observe that the visibility of the HOM dip decreases with higher losses and eventually, the dip even flips into a peak, indicating that indistinguishable photons are more likely to anti-bunch. Thus, the introduction of loss in a coupler directly affects the quantum interference of photons.

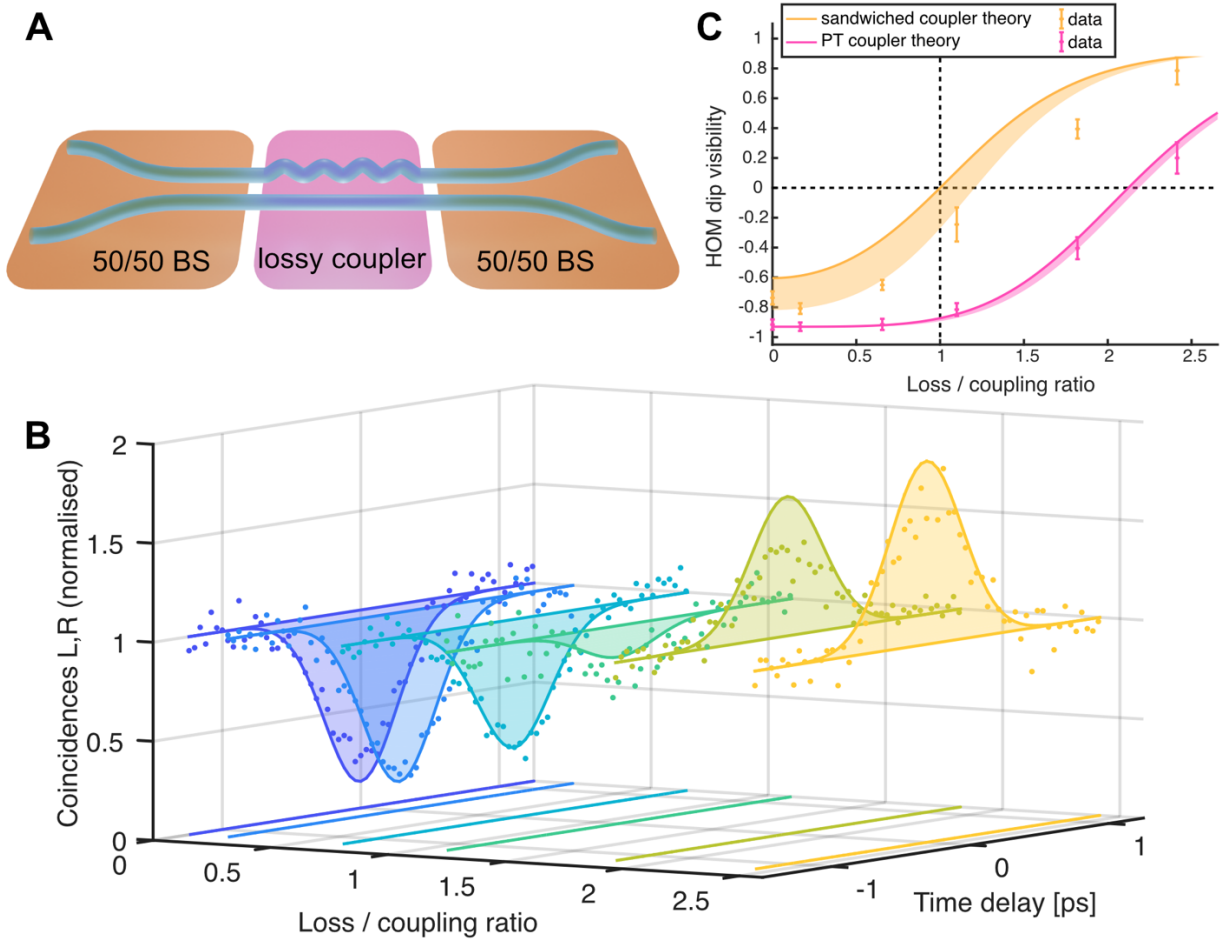

**Fig. 4. Direct signature of PT-symmetry breaking point in Hong-Ou-Mandel interference.** *(A) The lossy directional couplers are sandwiched between a pair of directional couplers of 50/50 beam splitting ratio. The samples vary in losses, corresponding to the structures in Fig. 3. (B) Hong-Ou Mandel interference is recorded by varying the time delay between photons. At the PT-symmetry breaking point, the dip is transformed into a peak. (C) The HOM dip visibilities for the single PT coupler and sandwiched system vary with loss to coupling ratio. For the sandwiched coupler, the HOM dip flips into a peak exactly at the EP. Error bars are based on data extracted from the HOM measurements in 4B) and 3E). The analytical theory for the experimental parameters is indicated by the lines, shaded areas represent expected deviations due to fabrication imperfections.*

**Indications of PT-symmetry breaking in two-photon correlations**

While the results so far show that the introduction of loss in the coupler alters the quantum behavior of photons, no direct signature of the exceptional point is visible. To demonstrate a prominent feature of the exceptional point on the quantum state evolving in the lossy coupler, we use a structure proposed by Longhi in (*31*). Here, the lossy coupler is sandwiched between a pair of 50/50 directional couplers that perform a forward and backwards rotation of the input/output state (see Fig. 4A) to change observation basis. With these rotations $R$, the resulting transformation of the sandwiched system is

$$U_{sandwich} = RUR^{-1}.$$

Notably, this configuration observes the lossy coupler in a basis that contains the single eigenvector of the lossy coupler at the exceptional point. We can express the dynamics of this rotated system using a Hamiltonian with asymmetric couplings and symmetric loss (see Supplementary Material), that reads

$$H_{sandwich} = \begin{pmatrix} -i\gamma & \kappa - \gamma \\ \kappa + \gamma & -i\gamma \end{pmatrix}$$

In this representation, the coupling from the second mode to the first becomes zero at the exceptional point, and its phase changes sign above the EP. Therefore, indistinguishable photon pairs being observed in an anti-bunched $|1,1\rangle$ output will see destructive interference below the exceptional point and constructive interference above, independent of the length of the coupler. Thus, the sandwiched system will exhibit a HOM dip in the unbroken PT-symmetric case that is transformed into a peak directly at the EP when PT-symmetry is broken (*32*).

In our experiment, we realize this structure by placing two lossless couplers of 50/50 splitting ratio before and after the lossy coupler. All parameters of the lossy coupler section are identical to the couplers in the previous experiment. We then perform a Hong-Ou-Mandel experiment on each of these sandwiched couplers, resulting in the data plotted in Fig. 4B). Here, the visibility of the HOM dip decreases with increasing loss, and we observe a pronounced peak in the last two samples that are in the broken PT symmetry phase. Figure 4C compares the measured visibilities of the HOM interference to the analytical theory for the lossy coupler and the sandwiched lossy directional coupler throughout the unbroken and broken PT-symmetry phase. The shaded areas represent expected deviations due to fabrication imperfections, such as coupler lengths. Note that, for the isolated lossy coupler, the point where a peak emerges is not tied to a specific loss/coupling ratio and depends for instance on the length of the coupler. In contrast, for the sandwiched coupler, the flip of the HOM dip clearly occurs at the exceptional point. In other words, the sign of the visibility indicates whether the system is in its PT broken or unbroken phase. More generally, our observations show that, while not directly obvious in a single lossy coupler, the EP does indeed significantly alter quantum interference of photons.

**Discussion**

In summary, we experimentally observe the impact of an exceptional point in a non-Hermitian coupler on the quantum interference of indistinguishable photons. With increasing loss, the probability of photons bunching into a single output decreases, and the probability of finding both photons in the lossless waveguide is systematically higher than the lossy waveguide, which can be linked to the photons avoiding loss (*40*). Investigating Hong-Ou-Mandel interference in the anti-bunching output, we observe that the dip in coincidences eventually transforms into a peak at high losses, confirming a theoretical prediction in (*32*). Surprisingly, these observables behave in a smooth way (*41*), with no distinct mark of the exceptional point. To demonstrate a direct signature of the EP in the two-photon correlations, the lossy couplers are sandwiched between a pair of 50/50 beam splitters. In this setup, the quantum interference of photon pairs changes sign at the exceptional point, resulting in destructive interference of anti-bunched photons below the EP (HOM dip) and constructive interference above, embodied as peak in coincidences.

Our work reveals a link between quantum correlations and exceptional points, and enables the study of quantum states in the broken PT-symmetry regime. Whether other signatures of the exceptional point might appear in quantum interference, how they translate to EPs of a higher order, or even extend to multiphoton interference, are intriguing open questions. Our findings shed new light onto the ways to dynamically cross an exceptional point and open new ways to study how information is lost there (*42*), a trajectory that is numerically hard to access. By exploiting an additional degree of freedom in a two-dimensional eigenvalue spectrum, one could dynamically encircle the exceptional point (*43*), flipping the quantum state of light from one eigenstate to the other which may be harnessed in quantum communication. Combining the enhanced sensitivity of exceptional-point-based sensors with quantum light might benefit quantum metrology (*44*). The interplay of quantum states with exceptional points remains largely unexplored, but is now within reach through our experimental platform.

## Materials and Methods

### Experimental Design

We fabricate our waveguide pairs using the femtosecond direct laser writing technique in fused silica (Corning 7980). A commercial laser system (Monaco, Coherent) supplies ultrashort laser pulses of 270 fs at a wavelength of 517 nm at a repetition rate of 333 kHz. The pulses are focused into the wafers using a x50 microscope objective with a numerical aperture of 0.6. By moving the sample under the laser using a three-axis motorized translation stage of 50 nm precision (Aerotech ALS180) the waveguides are formed in its volume (*45*).

In the beginning of our structure, the two waveguides are separated by 82 μm, which results in virtually zero coupling between the modes. A fan-in section draws the waveguides together, bringing them close enough to couple at a distance of 27 μm, and an identical fan-out structure follows at the output. In the lossy directional coupler section, we implement loss in one of the waveguides through rapid periodic bending. The lossy waveguide follows a sinusoidal path, where amplitude and period length of the sine control the loss in the waveguide (*39*, *46*). To realize different points in the PT-symmetry phase, the loss to coupling ratio needs to be varied. In our experiment, we start out with a lossless directional coupler of 21 mm coupling length, that acts as a 50/50 beam splitter. In the following samples, we keep coupling and the period length of the sine constant at $\kappa = 0.26\ cm^{-1}$ and $L_p = 3\ mm$, but increase the amplitude of the sine in 0.5 μm steps up to 3.5 μm. A separate characterization of the sinusoidal waveguides provides loss parameters in the range of $\gamma = 0\ ...\ 0.63\ cm^{-1}$.

### Measurement of two-photon Correlations

For the quantum experiment, two indistinguishable photons are generated through type I spontaneous parametric down conversion in a bismuth triborate (BiBO) crystal and collected into fibres. Changing the collection position of the photons allows to tune their time delay, and respectively their distinguishability. Through a fibre array with 82 μm pitch, the photons couple into the waveguide chip and propagate through the sample, the outcoupling is mediated by a second fibre array. We detect photon pairs using avalanche photo diodes (Excelitas) and a correlation card (Becker-Hickl). To access the full correlation matrix at the output, each output fibre serves as input into an additional 50/50 fibre beam splitter, resulting in four output signals, that are measured with four avalanche photo diodes.

# References

1. T. Kato, "Perturbation theory in a finite-dimensional space" in *Perturbation Theory for Linear Operators*, T. Kato, Ed. (Springer, Berlin, Heidelberg, 1966), pp. 62–126.

2. R. El-Ganainy, K. G. Makris, M. Khajavikhan, Z. H. Musslimani, S. Rotter, D. N. Christodoulides, Non-Hermitian physics and PT symmetry. *Nat. Phys.* **14**, 11–19 (2018).

3. N. Moiseyev, P. R. Certain, F. Weinhold, Resonance properties of complex-rotated hamiltonians. *Mol. Phys.* **36**, 1613–1630 (1978).

4. C. M. Bender, S. Boettcher, Real Spectra in Non-Hermitian Hamiltonians Having PT Symmetry. *Phys Rev Lett* **80**, 5243–5246 (1998).

5. K. G. Makris, R. El-Ganainy, D. N. Christodoulides, Z. H. Musslimani, Beam Dynamics in P T Symmetric Optical Lattices. *Phys. Rev. Lett.* **100**, 103904 (2008).

6. L. Feng, R. El-Ganainy, L. Ge, Non-Hermitian photonics based on parity–time symmetry. *Nat. Photonics* **11**, 752–762 (2017).

7. A. Li, H. Wei, M. Cotrufo, W. Chen, S. Mann, X. Ni, B. Xu, J. Chen, J. Wang, S. Fan, C.-W. Qiu, A. Alù, L. Chen, Exceptional points and non-Hermitian photonics at the nanoscale. *Nat. Nanotechnol.* **18**, 706–720 (2023).

8. C. Dembowski, H.-D. Gräf, H. L. Harney, A. Heine, W. D. Heiss, H. Rehfeld, A. Richter, Experimental Observation of the Topological Structure of Exceptional Points. *Phys. Rev. Lett.* **86**, 787–790 (2001).

9. A. Guo, G. J. Salamo, D. Duchesne, R. Morandotti, M. Volatier-Ravat, V. Aimez, G. A. Siviloglou, D. N. Christodoulides, Observation of PT-Symmetry Breaking in Complex Optical Potentials. *Phys. Rev. Lett.* **103**, 093902 (2009).

10. A. V. Sadovnikov, A. A. Zyablovsky, A. V. Dorofeenko, S. A. Nikitov, Exceptional-Point Phase Transition in Coupled Magnonic Waveguides. *Phys. Rev. Appl.* **18**, 024073 (2022).

11. Q. Zhong, R. El-Ganainy, Crossing exceptional points without phase transition. *Sci. Rep.* **9**, 134 (2019).

12. C. E. Rüter, K. G. Makris, R. El-Ganainy, D. N. Christodoulides, M. Segev, D. Kip, Observation of parity-time symmetry in optics. *Nat. Phys.*, doi: 10.1038/nphys1515 (2010).

13. J. Doppler, A. A. Mailybaev, J. Böhm, U. Kuhl, A. Girschik, F. Libisch, T. J. Milburn, P. Rabl, N. Moiseyev, S. Rotter, Dynamically encircling an exceptional point for asymmetric mode switching. *Nature* **537**, 76–79 (2016).

14. C. Guria, Q. Zhong, S. K. Ozdemir, Y. S. S. Patil, R. El-Ganainy, J. G. E. Harris, Resolving the topology of encircling multiple exceptional points. *Nat. Commun.* **15**, 1369 (2024).

15. H. Hodaei, A. U. Hassan, S. Wittek, H. Garcia-Gracia, R. El-Ganainy, D. N. Christodoulides, M. Khajavikhan, Enhanced sensitivity at higher-order exceptional points. *Nature* **548**, 187–191 (2017).

16. W. Chen, Ş. Kaya Özdemir, G. Zhao, J. Wiersig, L. Yang, Exceptional points enhance sensing in an optical microcavity. *Nature* **548**, 192–196 (2017).

17. H. Hodaei, M.-A. Miri, M. Heinrich, D. N. Christodoulides, M. Khajavikhan, Parity-time–symmetric microring lasers. *Science* **346**, 975–978 (2014).

18. L. Feng, Z. J. Wong, R.-M. Ma, Y. Wang, X. Zhang, Single-mode laser by parity-time symmetry breaking. *Science* **346**, 972–975 (2014).

19. H. Defienne, M. Barbieri, I. A. Walmsley, B. J. Smith, S. Gigan, Two-photon quantum walk in a multimode fiber. *Sci. Adv.* **2**, e1501054 (2016).

20. T. A. W. Wolterink, R. Uppu, G. Ctistis, W. L. Vos, K.-J. Boller, P. W. H. Pinkse, Programmable two-photon quantum interference in $10^3$ channels in opaque scattering media. *Phys. Rev. A* **93**, 053817 (2016).

21. B. Vest, M.-C. Dheur, É. Devaux, A. Baron, E. Rousseau, J.-P. Hugonin, J.-J. Greffet, G. Messin, F. Marquier, Anti-coalescence of bosons on a lossy beam splitter. *Science* **356**, 1373–1376 (2017).

22. S. M. Barnett, J. Jeffers, A. Gatti, R. Loudon, Quantum optics of lossy beam splitters. *Phys Rev A* **57**, 2134–2145 (1998).

23. R. Uppu, T. A. W. Wolterink, T. B. H. Tentrup, P. W. H. Pinkse, Quantum optics of lossy asymmetric beam splitters. *Opt Express* **24**, 16440–16449 (2016).

24. J. Huber, P. Kirton, S. Rotter, P. Rabl, Emergence of PT-symmetry breaking in open quantum systems. *SciPost Phys.* **9**, 052 (2020).

25. M. Znojil, Passage through exceptional point: case study. *Proc. R. Soc. Math. Phys. Eng. Sci.* **476**, 20190831 (2020).

26. W. Cao, X. Lu, X. Meng, J. Sun, H. Shen, Y. Xiao, Reservoir-Mediated Quantum Correlations in Non-Hermitian Optical System. *Phys. Rev. Lett.* **124**, 030401 (2020).

27. C. A. Downing, A. Vidiella-Barranco, Parametrically driving a quantum oscillator into exceptionality. *Sci. Rep.* **13**, 11004 (2023).

28. F. Roccati, A. Purkayastha, G. M. Palma, F. Ciccarello, Quantum correlations in dissipative gain–loss systems across exceptional points. *Eur. Phys. J. Spec. Top.* **232**, 1783–1788 (2023).

29. F. Klauck, L. Teuber, M. Ornigotti, M. Heinrich, S. Scheel, A. Szameit, Observation of PT-symmetric quantum interference. *Nat. Photonics* **13**, 883–887 (2019).

30. N. Maraviglia, P. Yard, R. Wakefield, J. Carolan, C. Sparrow, L. Chakhmakhchyan, C. Harrold, T. Hashimoto, N. Matsuda, A. K. Harter, Y. N. Joglekar, A. Laing, Photonic quantum simulations of coupled PT-symmetric Hamiltonians. *Phys. Rev. Res.* **4**, 013051 (2022).

31. S. Longhi, Quantum statistical signature of PT symmetry breaking. *Opt. Lett.* **45**, 1591–1594 (2020).

32. T. A. W. Wolterink, M. Heinrich, S. Scheel, A. Szameit, Order-Invariant Two-Photon Quantum Correlations in PT-Symmetric Interferometers. *ACS Photonics* **10**, 3451–3457 (2023).

33. C. K. Hong, Z. Y. Ou, L. Mandel, Measurement of subpicosecond time intervals between two photons by interference. *Phys Rev Lett* **59**, 2044–2046 (1987).

34. M. Ornigotti, A. Szameit, Quasi PT -symmetry in passive photonic lattices. *J. Opt.* **16**, 065501 (2014).

35. S. Scheel, A. Szameit, PT -symmetric photonic quantum systems with gain and loss do not exist. *EPL Europhys. Lett.* **122**, 34001 (2018).

36. L. Teuber, S. Scheel, Solving the quantum master equation of coupled harmonic oscillators with Lie-algebra methods. *Phys. Rev. A* **101**, 042124 (2020).

37. M. Gräfe, R. Heilmann, R. Keil, T. Eichelkraut, M. Heinrich, S. Nolte, A. Szameit, Correlations of indistinguishable particles in non-Hermitian lattices. *New J. Phys.* **15**, 033008 (2013).

38. A. Szameit, F. Dreisow, T. Pertsch, S. Nolte, A. Tünnermann, Control of directional evanescent coupling in fs laser written waveguides. *Opt Express* **15**, 1579–1587 (2007).

39. T. Eichelkraut, S. Weimann, S. Stützer, S. Nolte, A. Szameit, Radiation-loss management in modulated waveguides. *Opt Lett* **39**, 6831–6834 (2014).

40. M. Ehrhardt, M. Heinrich, A. Szameit, Observation-dependent suppression and enhancement of two-photon coincidences by tailored losses. *Nat. Photonics* **16**, 191–195 (2022).

41. S. Longhi, Quantum interference and exceptional points. *Opt. Lett.* **43**, 5371–5374 (2018).

42. B. Longstaff, E.-M. Graefe, Nonadiabatic transitions through exceptional points in the band structure of a PT-symmetric lattice. *Phys. Rev. A* **100**, 052119 (2019).

43. Z.-N. Tian, F. Yu, X.-L. Zhang, K. M. Lau, L.-C. Wang, J. Li, C. T. Chan, Q.-D. Chen, On-chip single-photon chirality encircling exceptional points. *Chip* **2**, 100066 (2023).

44. W. C. Wong, J. Li, Exceptional-point sensing with a quantum interferometer. *New J. Phys.* **25**, 033018 (2023).

45. K. M. Davis, K. Miura, N. Sugimoto, K. Hirao, Writing waveguides in glass with a femtosecond laser. *Opt Lett* **21**, 1729–1731 (1996).

46. S. Weimann, T. Eichelkraut, A. Szameit, Decay of bound states in oscillating potential wells. *Phys. Rev. A* **97**, 053844 (2018).

**Acknowledgments**

We thank C. Otto for preparing the high-quality fused silica samples used for the inscription of all photonic structures employed in this work.

**Funding:** AS acknowledges funding from the Deutsche Forschungsgemeinschaft (grants SZ 276/9-2, SZ 276/19-1, SZ 276/20-1, SZ 276/21-1, SZ 276/27-1, and GRK 2676/1-2023 'Imaging of Quantum Systems', project no. 437567992). AS also acknowledges funding from the Krupp von Bohlen and Halbach Foundation as well as from the FET Open Grant EPIQUS (grant no. 899368) within the framework of the European H2020 programme for Excellent Science. AS and MH acknowledge funding from the Deutsche Forschungsgemeinschaft via SFB 1477 'Light–Matter Interactions at Interfaces' (project no. 441234705). T.A.W.W. is supported by a European Commission Marie Skłodowska-Curie Actions Individual Fellowship "Quantum correlations in PT-symmetric photonic integrated circuits", Project No. 895254.

**Author contributions:**
Conceptualization: FK, AS, TAWW
Methodology: FK, TAWW
Investigation: FK
Visualization: FK
Supervision: TAWW
Writing—original draft: FK
Writing—review & editing: FK, MH, AS, TAWW

**Competing interests:** The authors declare they have no competing interests.

**Data and materials availability:** All data are available in the main text or the supplementary materials.

**Supplementary Materials**

Hamiltonian of the sandwiched coupler

The lossy directional coupler is sandwiched between two 50/50 couplers, acting as rotation matrices. The time evolution can be described by

$$U_{sandwich} = R\, U R^{-1} = R e^{iHz} R^{-1} = e^{iRHR^{-1}z} = e^{iH_{sandwich}z}$$

The Hamiltonian of the sandwiched coupler is therefore given as:

$$H_{sandwich} = R\, H R^{-1}$$

$$= \frac{1}{\sqrt{2}}\begin{pmatrix} 1 & -i \\ -i & 1 \end{pmatrix}\begin{pmatrix} 0 & \kappa \\ \kappa & -2i\gamma \end{pmatrix}\frac{1}{\sqrt{2}}\begin{pmatrix} 1 & i \\ i & 1 \end{pmatrix}$$

$$= \frac{1}{2}\begin{pmatrix} 1 & -i \\ -i & 1 \end{pmatrix}\begin{pmatrix} i\kappa & \kappa \\ \kappa + 2\gamma & i\kappa - 2i\gamma \end{pmatrix}$$

$$= \begin{pmatrix} -i\gamma & \kappa - \gamma \\ \kappa + \gamma & -i\gamma \end{pmatrix}$$